\documentclass[11pt,a4paper,english,superscriptaddress,aps,preprint]{revtex4}
\usepackage{amsmath}
\usepackage{amssymb}
\makeatletter
\newcommand{\bea}{\begin{eqnarray}}
\newcommand{\eea}{\end{eqnarray}}

\newcommand{\e}{\varepsilon}
\newcommand{\pa}{\partial}
\newcommand{\be}{\begin{equation}}
\newcommand{\ee}{\end{equation}}
\newcommand{\bq}{\begin{eqnarray}}
\newcommand{\eq}{\end{eqnarray}}

\begin{document}

\title{On the duality in CPT-even Lorentz-breaking theories}
\author{A. P. Ba\^eta Scarpelli}
\affiliation{Setor T\'{e}cnico-Cient\'{\i}fico - Departamento de Pol\'\i cia Federal
Rua Hugo D'Antola, 95 - Lapa - S\~{a}o Paulo - Brazil}
\email{scarpelli.apbs@dpf.gov.br}
\author{R. F. Ribeiro}
\affiliation{Departamento de F\'{\i}sica, Universidade Federal da Para\'{\i}ba\\
 Caixa Postal 5008, 58051-970, Jo\~ao Pessoa, Para\'{\i}ba, Brazil}
\email{rfreire, jroberto, petrov@fisica.ufpb.br}
\author{J. R. Nascimento}
\affiliation{Departamento de F\'{\i}sica, Universidade Federal da Para\'{\i}ba\\
 Caixa Postal 5008, 58051-970, Jo\~ao Pessoa, Para\'{\i}ba, Brazil}
\email{rfreire, jroberto, petrov@fisica.ufpb.br}
\author{A. Yu. Petrov}
\affiliation{Departamento de F\'{\i}sica, Universidade Federal da Para\'{\i}ba\\
 Caixa Postal 5008, 58051-970, Jo\~ao Pessoa, Para\'{\i}ba, Brazil}
\email{rfreire, jroberto, petrov@fisica.ufpb.br}

\begin{abstract}
We generalize the duality between self-dual and Maxwell-Chern-Simons theories for the case of a CPT-even Lorentz-breaking extension of these theories. The duality is shown with use of the gauge embedding procedure, both in free and coupled cases, and with the master action approach. The physical spectra of both Lorentz-breaking theories are studied. The massive poles are shown to coincide and to respect the requirements for unitarity and causality at tree level. The extra massless poles which are present in the dualized model are shown to be nondynamical.
\end{abstract}

\maketitle

\section{Introduction}

Lorentz-breaking theories have attracted great attention in the last two decades (for a general review on this issue, see \cite{Kostel} and references therein). The most generic Lorentz-breaking extension of the free 4-dimensional gauge vector field theory discussed in \cite{Kostel} looks like
\bea
{\cal L}_{vect}=-\frac{1}{4}F_{mn}F^{mn}-\frac{1}{4}(k_F)_{mnpq}F^{mn}F^{pq}+\frac{1}{2}(k_{AF})^m\epsilon_{mnpq}A^nF^{pq}-(k_A)_m A^m.
\eea
If we reduce our study to 3-dimensional space-time, we should replace the Carroll-Field-Jackiw (CFJ) term, $(k_{AF})^m\epsilon_{mnpq}A^nF^{pq}$, with the Chern-Simons one, $m\epsilon_{npq}A^nF^{pq}$. Also, we can disregard the linear term, $(k_A)_mA^m$, since it does not propagate and yields a trivial contribution at the quantum level. So, we rest with
\bea
 {\cal L}_{vect}=-\frac{1}{4}F_{mn}F^{mn}-\frac{1}{4}(k_F)_{mnpq}F^{mn}F^{pq}+\frac{1}{2}m\epsilon_{npq}A^nF^{pq}.
\eea
Just this theory will be obtained in this paper through the CPT-even Lorentz-breaking extension of the self-dual theory, whose lagrangian density is given by
\bea
  {\cal L}_{SD}=\frac{m^2}{2}A^m(\eta_{mn}+\kappa_{mn})A^n+\frac{1}{2}m\epsilon_{npq}A^nF^{pq},
\eea
in which $\kappa_{mn}$ is a constant symmetric tensor. We note that an analogous study for the CPT-odd Lorentz-breaking extension of the self-dual theory has been carried out in \cite{Fur}.

The concept of duality between two different models in field theory is very interesting and useful, allowing for mutual mapping of theories possessing essentially different actions, since there are some important features which are manifest in one model but are hidden in the other one. Besides of this, the duality allows to map the weak-coupled theory to the strong-coupled one because of the implementation of the relation between electric and magnetic couplings. Duality was first established in three space-time dimensions in the paradigmatic example of the dual correspondence between the free self-dual and Maxwell-Chern-Simons theories \cite{DJ,Nieuw} and has been discussed as a generic feature of a wide class of field theory models in \cite{Nieuw}.

Since then, different methods have been elaborated to establish and study the duality in many cases (see \cite{review} for a nice review). One powerful approach to determine the physical equivalence between two theories is the master action method \cite{Malacarne}, whose essence consists in determining an action involving two vector fields. The two models can be obtained from the master action by using the equations of motion of the fields in the original action. On the other hand, the gauge embedding method \cite{Anacleto} is based on the transformation of the self-dual model in a gauge theory by adding on mass shell  vanishing terms. This approach, accomplished by an iterative embedding of Noether counterterms, is based on the idea of a local lifting of a global symmetry and is reminiscent to the papers by Freedman and van Nieuwenhuizen \cite{Nieuw2} and subsequent works by Ferrara, Freedman and van Nieuwenhuizen \cite{Ferrara1} and Ferrara and Scherk \cite{Ferrara2}, which were important for the construction of component-field supergravity actions.

These methods have been shown to be efficient tools for studying different field theory models, allowing, in particular, to find new couplings for vector fields. As a good example, the self-dual theory minimally coupled to the spinor matter has been shown to generate, through gauge embedding, the magnetic (nonminimal) and the Thirring-like current-current couplings \cite{Anacleto}. Further, the duality has been established between nonlinear generalizations of self-dual and Maxwell-Chern-Simons theories (the last one yields a Born-Infeld-Chern-Simons theory) \cite{nlin}, higher-rank tensor generalizations of these theories \cite{highrank} and their higher-derivative extensions \cite{HD}. Noncommutative extensions of the duality have been discussed in \cite{NC}.

Several papers have been dedicated to the extension of duality to Lorentz-breaking models in recent years. The duality methodology has been applied to CPT-odd Lorentz-breaking models, like in the extension of the 3D self-dual theory \cite{Fur} and in its promotion to four-dimensional space-time \cite{Bota}, \cite{Wo}. The Standard Model Extension (SME) \cite{kostelecky1}-\cite{coleman2}, which provides a description of Lorentz and CPT violation in Quantum Field Theories, also includes CPT-even terms. These CPT-even relativity-breaking models have been the focus of intense investigation recently and many issues related to classical solutions in these theories have been discussed  (a very incomplete list is given in \cite{aethercl,aether}). However, the dualization of CPT-even Lorentz-violating models has not been given much attention. Recently, the dual embedding of a four-dimensional Proca-like theory with a CPT-even Lorentz-breaking mass term was carried out \cite{Scarp}. The resulting theory was shown to involve very  interesting higher derivatives terms.

We are particularly interested in the investigation of $3D$ dual CPT-even Lorentz-breaking models. First of all, the study of $3D$ models is very instigating. Planar physics (in $2+1$ dimensions) presents many interesting surprises, both experimentally and theoretically, since the behavior of fermion and gauge fields differs from what we are used to in classical and quantum electrodynamics. For example, Chern-Simons theories are interesting both for their theoretical novelty, and for their practical application in planar condensed matter phenomena, such as the fractional quantum Hall effect (see, for example, \cite{Dunne}). Particularly in $2+1$ dimensions, Lorentz-violating models find a branch of applications. It is surprising that a CPT-odd Lorentz-breaking effective Lagrangian emerges from a full microscopic model for Weyl semi-metals \cite{Grushin}. Another good example is the one of a $3D$ relativity-breaking model with four-fermion interactions \cite{JRN}. It is shown that its low-energy limit encompasses a branch of sub-models which resemble those used in the study of graphene.

In this paper, we construct a CPT-even Lorentz-breaking generalization of the famous 3D duality between self-dual and Maxwell-Chern-Simons theories of \cite{DJ}. For this, we employ the gauge embedding method and, further, we check the duality through different methods, both in free and coupled cases.

The paper is organized as follows: section II is dedicated to the presentation of the self-dual (SD) model, the determination of the corresponding dual Maxwell-Chern-Simons-like (MCS) theory by means of the gauge embedding technique and the confirmation of duality through the analysis of the equations of motion; section III is devoted to the confirmation of this duality with the use of the master action formalism; in section IV, we study the physical consistency of both models through the analysis of their spectra, obtained from the propagators. The physical equivalence of the models is also shown from the physical spectra viewpoint; the concluding remarks are presented in section V. Some technical details are left for the appendix in section VI.

\section{Gauge embedding}

Let us consider the following generalization of the $3D$ self-dual model \cite{DJ},
\bea
\label{sd1}
{\cal L}=-\frac{m}{2}\varepsilon^{anb}f_a\pa_n f_b+\frac{1}{2}m^2f^ah_{ab}f^b+f_aj^a,
\eea
where we added an interaction term (the current $j_a$ can be, for example, the spinorial one, $j^a=\bar{\psi}\gamma^a\psi$) and $h_{ab}$ is a tensor which includes Lorentz-violating terms.
An interesting question which emerges is whether it is possible to obtain, from the model of eq. (\ref{sd1}), a gauge invariant physical equivalent theory. We proceed to the gauge embedding of our model. The method consists in a two-step Noether embedding of the gauge symmetry $\delta f_a=\partial_{a}\eta$ of ${\cal L}$ without the mass term. For this, it is used an auxiliary field $B_a$, such that $\delta B_a=\delta f_a=\partial_{a}\eta$, in order to restore gauge symmetry. Let us then calculate the first variation of our Lagrangian density,
\be
\delta {\cal L}[f_a]\,=
\left\{-m\varepsilon^{abc}\partial_b f_c + m^2 h^{ab} f_b + j^a\right\} \delta f_a,
\ee
in which we recognize the Noether current as
\be
\label{e2}
K^a=-m\varepsilon^{abc}\partial_b f_c + m^2 h^{ab} f_b + j^a.
\ee
The first iterated Lagrangian is constructed by introducing the auxiliary field $B$,
\be
{\cal L}^{(1)}\,=\,{\cal L}\,-\,K^a B_a,
\ee
with $\delta B_{a}\,=\,\delta f_{a} = \partial_{a} \eta$, so that we get
\be
\delta\,{\cal L}^{(1)}\,=\,-(\delta\,K_{a})\,B^{a}.
\ee
Using
\be
\delta K^a=m^2 h^{ab} \delta f_b,
\ee
we have
\be
\label{varl1}
\delta {\cal L}^{(1)}=-m^2 B_a h^{ab} \delta f_b.
\ee
The second iterated Lagrangian is defined by
\be
{\cal L}^{(2)}\,=\,{\cal L}^{(1)}\,+\,\frac{m^2}{2}\,B^{a}h_{ab}B^b,
\ee
so that if we use the variation of $B_a$ and (\ref{varl1}), we get that the total variation vanishes, $\delta {\cal L}^{(2)}\,=\,0$. Let us write down the
explicit form of this action,
\bq
\label{l2}
&&{\cal L}^{(2)}= -\frac 12 m \varepsilon^{abc}f_a \partial_b f_c  + \frac {m^2}{2}f^a h_{ab} f^b + f_aj^a \nonumber \\
&&- K_a B^a + \frac{m^2}{2}\,B^{a}h_{ab}B^b.
\eq
After carrying out the variation of this action with respect to $B_a$, we get the equation of motion,
\be
K_a-m^2h_{ab} B^b=0,
\ee
which plugged back into (\ref{l2}), will give us the gauge invariant Maxwell-Chern-Simons-like action,
\bea
{\cal L}_{MCS}=\frac{m}{2}F^aA_a-\frac{1}{2}F^a(h^{-1})_{ad}F^d-\frac{1}{2m^2}G^a(h^{-1})_{ad}G^d+\frac{1}{m}F^a(h^{-1})_{ad}G^d,
\label{mcs1}
\eea
in which we renamed the field $f$ as $A$ and the current $j$ as $G$ and used $F^a=\e^{abc}\pa_b A_c$.

In this paper we will concentrate on the specific case where the Lorentz-violating tensor is given by $h_{ad}=\eta_{ad}-\beta b_a b_d$, where $b_a$ is a Lorentz-breaking constant background field which selects a preferred direction in the $3D$ space-time and $\beta$ is a dimensionless parameter. So, we have
\be
(h^{-1})_{ad}=\eta_{ad}+\alpha b_a b_d,
\ee
with $\alpha=\beta/(1-\beta b^2)$. The MCS-like Lagrangian density is then given by
\bq
&&{\cal L}_{MCS}=-\frac{1}{4}F_{ab}F^{ab}+\frac{m}{2}\e^{abc}A_a\pa_b A_c-\frac{\alpha}{8}\left(\e_{abc}b^aF^{bc}\right)^2 \nonumber \\
&& -\frac{1}{2m^2}G^a(h^{-1})_{ad}G^d+\frac{1}{m}F^a(h^{-1})_{ad}G^d.
\eq
It is interesting to note that the quadratic term in the Levi-Civita symbol, which emerged from the gauge embedding procedure, is nothing but a linear combination of a Maxwell term and the aether term of \cite{Carroll}, \cite{aether}:
\bea
\left(\epsilon_{abc}b^aF^{bc}\right)^2=2b^2F_{ab}F^{ab}-4b_aF^{ac}b^bF_{bc}.
\eea
Besides, we note that this Lagrangian density involves a Thirring-like current-current interaction and a magnetic coupling, just like in the Lorentz-invariant case \cite{Anacleto}.

The next step towards the checking of the physical equivalence of the models is the analysis of the field equations. After some simple algebraic manipulations, the equations of motion for the self-dual and MCS fields, respectively, will look like
\be
mf_k- \e^{nlm}(h^{-1})_{kn}\pa_lf_m=-\frac{1}{m}j^n(h^{-1})_{kn}
\ee
and
\be
mF^n-\e^{man}(h^{-1})_{ab}\pa_mF^b=-\frac{1}{m}\e^{lmn}(h^{-1})_{md}\pa_l G^d,
\ee
or, in terms of the fields $f_n$ and $\tilde{F}_p=F^b(h^{-1})_{bp}$,
\be
\left[mh^{bn}-\e^{bmn}\pa_m\right]f_n=-\frac{1}{m}j^b
\ee
and
\be
\left[mh^{bn}-\e^{bmn}\pa_m\right]\tilde{F}_n=-\frac{1}{m}\e^{abc}(h^{-1})_{cd}\pa_a G^d.
\ee
We see that the fields $f_n$ and $\tilde{F}_n$ satisfy similar equations. The mapping of currents $j^b\to \e^{abc}(h^{-1})_{cd}\pa_a G^d$ confirms the duality between our extended SD and MCS theories.
It is clear that in the Lorentz-invariant case, for which $h_{bk}=\eta_{bk}$, the known result from \cite{Anacleto} is reproduced.

\section{Master action approach}

It is interesting to establish the duality discussed in the last section in the framework of a master action. We would like to show that there is a consistent master action which generates the two actions studied in the gauge embedding approach. Indeed, let us consider the following Lagrangian density describing the dynamics of two vector fields, $f_a$ and $A_a$, and verify that it originates the free parts of the two dual models of the previous section,
\bea
\label{master}
{\cal L}_M=\frac{m^2}{2}f^ah_{ab}f^b+mf_a\epsilon^{abc}\pa_bA_c+\frac{m}{2}\epsilon^{abc}A_a\pa_bA_c+\frac{\rho}{2}(\pa\cdot A)^2,
\eea
in which the last part represents a gauge-fixing term. First, let us write down the equation of motion for $f_a$, which will be given by
\bea
f_a=-\frac{1}{m}(h^{-1})_a^{\phantom{a}n}\epsilon_{nbc}\pa^bA^c\equiv-\frac{1}{m}(h^{-1})_{an}F^n.
\eea
With the use of the above equation we can eliminate the field $f_a$  from the action (\ref{master}) and get
\bea
{\cal L}_A=-\frac{1}{2}F^a(h^{-1})_{ab}F^b+\frac{m}{2}F^a A_a+\frac{\rho}{2}(\pa\cdot A)^2,
\eea
which is nothing but the free Lagrangian density (\ref{mcs1}) with the gauge-fixing term.

The same procedure follows for the field $A_a$, for which the equation of motion reads
\bea
-m\e^{abc}\pa_bf_a-m\e^{abc}\pa_bA_a-\rho\pa^c(\pa\cdot A)=0,
\eea
so that
\bea
\label{aa}
A_a=m(\Delta^{-1})_{ac}\epsilon^{mbc}\pa_bf_m,
\eea
with
\bea
\Delta^{ac}=m\epsilon^{abc}\pa_b-\rho\pa^a\pa^c.
\eea
Inverting $\Delta^{ac}$,
\bea
(\Delta^{-1})_{ac}=-\frac{\pa_a\pa_c}{\rho\Box^2}-\frac{1}{m\Box}\epsilon_{acd}\pa^d,
\eea
and substituting this expression in (\ref{aa}), one finds
\bea
A_a=-f_a+\frac{\pa_a}{\Box}(\pa\cdot f),
\eea
which fixes $\pa\cdot A=0$. After the elimination of $A_a$ from (\ref{master}), we get
\bea
{\cal L}_f=\frac{m^2}{2}f^ah_{ab}f^b-\frac{m}{2}f_a\epsilon^{abc}\pa_bf_c,
\eea
which is the self-dual lagrangian density (\ref{sd1}). We thus have shown that (\ref{master}) is a master action (under integration) for our Lorentz-violating self-dual and Maxwell-Chern-Simons models.

\section{Propagators and structure of the poles}

In the previous sections we have established the duality between the Lorentz-violating self-dual (SD) and Maxwell-Chern-Simons (MCS) models. First we obtained the MCS model by means of the gauge embedding procedure and then compared the two equations of motion, finding a mapping between the two vector fields $f_a$ and $A_a$. In the sequence we have shown that there exists a master action which generates the two models. However, it is necessary a further investigation. It was observed in \cite{Bota} and shown in \cite{scarp-epl} that, although dual models share the same physical spectrum, the gauge invariant model obtained trough gauge embedding (also called Noether Dualization Method) exhibits new nonphysical poles. In this section we will study the propagators and show that the two models share the same physical spectrum and, besides, that the new poles which appear in the MCS model have no dynamics. It is known that Lorentz-violating models could have problems with stability and unitarity, as it was shown in the detailed discussion of \cite{Kost-Lehnert}. So, we also carry out in this section a study of the conditions under which these physical properties are preserved.

In the analysis below, we will consider only the quadratic part of the Lagrangian densities, which, under partial integration, are written in the form
\be
{\cal L}=\frac 12 u^a {\cal O}_{ab} u^b,
\ee
where $u_a$ represents the corresponding vector field. The propagator is given by $i \left({\cal O}^{-1}\right)_{ab}$. We will perform the calculations with the use of the following set of spin operators:
\bq
&&\theta_{ab}=\eta_{ab}-\frac{\partial_a \partial_b}{\Box}, \,\,\, \omega_{ab}= \frac{\partial_a \partial_b}{\Box}, \,\,\, S_{ab}=\varepsilon_{abc}\partial^c, \,\,\,
\Lambda_{ab}= b_a b_b,  \nonumber \\
&&\Sigma_{ab}=b_a \partial_b, \,\,\, A_{ab}=\tilde \Sigma_a b_b, \,\,\, B_{ab}= \tilde \Sigma_a \partial_b,
\eq
where $\tilde \Sigma_a = \varepsilon_{abc}\Sigma^{bc}$ ($\lambda$ stands for
$\Sigma _a \,\,^a=b_a \partial^a$), whose algebra is presented in the appendix.

\subsection{Self-dual model}

The quadratic part of the Lagrangian density of the self-dual model is given by
\be
\label{sd}
{\cal L}_f=\frac 12 f^a K_{ab} f^b,
\ee
with
\be
K_{ab}=m^2 \theta_{ab} + m^2 \omega_{ab} + m S_{ab} -\beta m^2 \Lambda_{ab}.
\ee
The inverse of this wave operator is obtained with the use of the algebra of table 1 of the appendix and is given by
\be
(K^{-1})_{ab} \equiv G_{ab}=\frac {1}{R+m^2}\left\{ \theta_{ab}+\frac{1}{m^2}\left(R+m^2+\alpha \lambda^2\right)\omega_{ab} - \frac 1m S_{ab} +\alpha \Lambda_{ab}
- \frac{\alpha}{m}\left(A_{ab}-A_{ba}\right)\right\},
\ee
where
\be
R=\left( 1+\alpha b^2\right)\left(\Box -\beta \lambda^2\right),
\ee
From the above propagator, we obtain the dispersion relation for our self-dual theory, which looks like
\bea
\label{disprel}
-E^2+\mathbf{p}^2+m^2-\alpha\left[b^2(E^2-\mathbf{p}^2)-(b_0E-\mathbf{b}\cdot\mathbf{p})^2\right]=0.
\eea

We are now in position to study this dispersion relation and the physical spectrum of the model. We are interested in two situations for the background vector $b_\mu$, namely the cases in which it is spacelike or timelike.

\subsubsection{$b_a$ spacelike}

We use a representative background vector given by $b^m=(0,0,t)$. In this case, we have
\be
b \cdot p=t p_2  \,\,\,\,\,\, \mbox{and} \,\,\,\,\,\, b^2=-t^2,
\label{spacelike}
\ee
and the dispersion relation yields
\be
E^2-\mathbf{p}^2-(1+\beta t^2)(m^2+\alpha t^2 p_2^2)=0.
\ee
It is interesting to check under which conditions this model, with $b^m$ spacelike, could yield a spacelike momentum $p^m$. First, let us right
\be
E^2-\mathbf{p}^2=(1+\beta t^2)\left(m^2+\frac{\beta t^2}{1+\beta t^2} p_2^2\right).
\ee
It is easy to see that if $\beta>0$, there is no possibility of $p^\mu$ being spacelike. On the other hand, if we set $\beta=-1$, we have
\be
E^2- \mathbf{p}^2=(1-t^2) m^2-t^2 p_2^2,
\ee
which will give us a spacelike momentum only in the case
\be
t^2>\frac{m^2}{m^2+p_2^2}.
\ee
We then conclude the model is stable {(that is, unitary)} for a little deviation from Lorentz symmetry, that is, in a concordant frame \cite{Kost-Lehnert}.

Concerning the microcausality, we will have supraluminal modes if $|\pa E /\pa p^m|>1$. For our spacelike $b^m$, we have
\be
\frac{\pa E}{\pa p^i}=\frac{p_i+\beta(\mathbf{b}\cdot \mathbf{p})b_i}{E}.
\ee
For a extreme situation, in which $\mathbf{p}$ and $\mathbf{b}$ are parallel, we have
\be
\frac{\pa E}{\pa p^i}=(1+\beta t^2)\frac{p_i}{E}
\ee
We observe that the satisfaction of microcausality depends on the magnitude of the Lorentz-breaking parameter, on the sign of $\beta$ and on the stability of the model.

Our present task consists in checking the features of the pole of the propagator for $b_a$ spacelike. In order to investigate the physical nature of the simple pole, we need to calculate the eigenvalues of the residue matrix of the propagator for this pole.  In this analysis, we are interested in checking the degrees of freedom of this mode and if it respects physical requests such as unitarity and causality.

For the choice we have made for $b^m$, the pole of the propagator is given by
\be
m_1^2=p_1^2+(1+\beta t^2) \left(m^2+p_2^2\right)
\ee
and the residue matrix reads
\be
{\cal R}=
\left(
\begin{array}{cccc}
\frac{(1+\beta t^2)p_2^2+ p_1^2}{m^2}& \frac{m_1 p_1-i (1+\beta t^2)p_2m}{m^2} & \frac{m_1 p_2 +ip_1m}{m^2}\\
\frac{m_1 p_1+i (1+\beta t^2)p_2m}{m^2} & \frac{m_1^2-(1+\beta t^2)p_2^2}{m^2} & \frac{p_1 p_2+ i m_1m}{m^2}  \\
\frac{m_1 p_2-ip_1m}{m^2} &  \frac{p_1 p_2-i m_1m}{m^2}  & \frac{m^2+p_2^2}{m^2}
\end{array}
\right),
\ee
with eigenvalues
\bq
\lambda_1 &=& 0,  \\
\lambda_2 &=& 0,  \\
\lambda_3 &=& \frac{1}{m^2}\left(m^2+m_1^2+\mathbf{p}^2\right).
\eq

As it can be seen, for $\beta>0$ we have one positive eigenvalue. For $\beta=-1$, if
\be
t^2<2\frac{m^2+\mathbf{p}^2}{m^2+p_2^2},
\ee
we have $\lambda_3>0$ and this pole is to be associated with one physical degree of freedom.

\subsubsection{$b_a$ timelike}

For $b_a$ timelike, we use a representative background vector given by $b^m=(t,0,0)$, which will give us
\be
b \cdot p=t p_0  \,\,\,\,\,\, \mbox{and} \,\,\,\,\,\, b^2=t^2,
\label{spacelike1}
\ee
and the dispersion relation,
\be
E^2-(1+\alpha t^2)\mathbf{p}^2 - m^2=0.
\ee
Since now we have $\alpha=\frac{\beta}{1-\beta t^2}$, we will have a spacelike momentum for $\beta<0$ if
\be
t^2>\frac{m^2}{m^2-\mathbf{p}^2},
\ee
with $\beta=-1$, for example. If $\beta>0$, we have problems with stability when
\be
1<t^2<\frac{m^2}{m^2-\mathbf{p}^2},
\ee
where we have set $\beta=1$. In this case, the expression only make sense when $m^2>\mathbf{p}^2$. It is clear, however, that the model is stable for a tiny Lorentz-breaking ($t^2<<1$), regardless the sign of $\beta$.

Considering the microcausality analysis, we obtain
\be
\frac{\pa E}{\pa p^i}=(1+\alpha t^2)\frac{p_i}{E},
\ee
which is similar to the one written for $b_m$ spacelike.

The propagator will yield the pole
\be
m_1'^2=(1+\alpha t^2)\mathbf{p}^2 + m^2,
\ee
for which the residue matrix will have only one nonnull eigenvalue, given by
\be
\lambda_3=1+ \frac{m_1'^2+(1+\alpha t^2)^2\mathbf{p}^2}{m^2}.
\ee
If $\beta<0$, we have $\lambda_3>0$. On the other hand, if, for example, $\beta=1$, and we write $m^2=\kappa \mathbf{p}^2$, we will have a positive eigenvalue for
\be
t^2>1+ \frac{1 +\sqrt{1-8\kappa}}{4\kappa} \,\,\,\mbox{or} \,\,\,
t^2<1-\frac{1-\sqrt{1-8\kappa}}{4\kappa},
\ee
It is important to note that, if $m^2>\mathbf{p}^2/8$, the nonnull eigenvalue is positive, complying with unitarity. The positivity of $\lambda_3$ can be preserved for all values of $t^2$, except $t^2=3$, depending on the value of $\kappa$ (or, more specifically, the relation between $m^2$ and $\mathbf{p}^2$). For $t^2<2$, $\lambda_3$ is always positive. We see that at tree level the model predicts a mode which complies with unitarity (positive norm particle) and causality (positive pole) for both spacelike and timelike $b_a$, as long as some conditions are imposed in the magnitude of the Lorentz-violating parameter. Specifically, for a little deviation from Lorentz symmetry, which is realized by $t^2<<1$, these physical properties are preserved.

\subsection{Maxwell-Chern-Simons model}

We now consider the quadratic part of the Lagrangian density of the MCS-like Lorentz-violating model. One can fix the gauge by adding the rescaled usual gauge-fixing term $-\frac{1}{2}(1+\alpha b^2)(\pa\cdot A)^2$. Afterwards, one gets after partial integrations,
\bea
{\cal L}_A=\frac{1}{2}A^a\Delta_{ac}A^c,
\eea
with
\bea
\label{delta}
\Delta_{ab}=R \theta_{ab} + R \omega_{ab} - m S_{ab} -\alpha \Box \Lambda_{ab} + \alpha \lambda \left(\Sigma_{ab}+\Sigma_{ba}\right).
\eea
The propagator is obtained with the inversion of this wave operator. With the help of the algebra of table 1 of the appendix, we get
\bq
&&(\Delta^{-1})_{ab} \equiv \tilde G_{ab}=\frac {1}{R+m^2}\left\{ \theta_{ab}+ \frac {1}{\Box}\left[\left(1-\beta b^2\right) (R+m^2) + \alpha \lambda^2\right]\omega_{ab} +\frac {m}{R}S_{ab} + \alpha \Lambda_{ab} \right. \nonumber \\
&&\left. -\frac {\alpha \lambda}{\Box} \left(\Sigma_{ab}+\Sigma_{ba}\right)+ \frac{\alpha m}{R}\left(A_{ab}-A_{ba}\right)-\frac {\alpha m \lambda}{\Box R}\left(B_{ab}-B_{ba}\right)\right\},
\eq
Before carrying out the same analysis which was performed for the self-dual model, with the study of the residues at the poles, let us remark that we have now, besides the pole $m_1^2$ found in the self-dual model, two more poles which appear in some sectors of the propagator, due to the factors $R$ and $\Box$ in the denominator. Let us argue that these poles are actually non-dynamical. If we saturate the propagator with conserved currents,
\be
{\cal SP} \equiv J^a i \tilde G_{ab} J^b,
\ee
such that $\pa_a J^a=0$ (or $p_a J^a=0$ in momentum space), we will have $J^a\omega_{ab}J^b=0$, $J^a(\Sigma_{ab}+\Sigma_{ba})J^b=0$ and $J^a(B_{ab}-B_{ba})J^b=0$, so that we rest with
\be
{\cal SP} = i J^a\frac {1}{R+m^2}\left\{ \theta_{ab}+\frac {m}{R}S_{ab} + \alpha \Lambda_{ab} +  \frac{\alpha m}{R}\left(A_{ab}-A_{ba}\right) \right\}J^b.
\ee
We remain with the terms involving $S_{ab}$ and $A_{ab}-A_{ba}$, proportional to $\frac{1}{R}$. However, they can be treated as analogues of the terms involving massless poles for the usual Maxwell-Chern-Simons theory. Indeed, in the Lorentz-invariant limit $b_a=0$, the usual Maxwell-Chern-Simons propagator is recovered. Those terms with massless poles are well known to yield no physical dynamics \cite{DJT}. In three-dimensional gauge theories, there is only one degree of freedom and, at the same time, it was noted in \cite{Fur} (see also the references therein) that in dual theories only physical dispersion relations should coincide. Therefore, despite the existing massless pole, we conclude that only the degree of freedom corresponding to the denominator $R+m^2$ is physical. Thus, we only have to check the residues for the pole $m_1^2$ corresponding just to this denominator.

\subsubsection{$b_a$ spacelike}

Adopting the same choice for $b_a$ we have used before ($b^a=(0,0,t)$), the residue in the pole $m_1^2$ will give us the matrix
\be
{\cal R}=
(1+\beta t^2)\left(
\begin{array}{cccc}
\frac{\rho m_1^2}{(m_1^2-\mathbf{p}^2)}-1& \frac{\rho m_1 p_1-i m p_2}{(m_1^2-\mathbf{p}^2)} & \frac{m_1 p_2 m^2+i p_1 m(m_1^2-\mathbf{p}^2)}{(m_1^2-\mathbf{p}^2)^2} \\
\frac{\rho m_1 p_1+i m p_2}{(m_1^2-\mathbf{p}^2)} & 1 + \frac{\rho p_1^2}{(m_1^2-\mathbf{p}^2)}&
\frac{p_1 p_2 m^2+i m_1 m(m_1^2-\mathbf{p}^2)}{(m_1^2-\mathbf{p}^2)^2} \\
\frac{m_1 p_2 m^2-i p_1 m(m_1^2-\mathbf{p}^2)}{(m_1^2-\mathbf{p}^2)^2} & \frac{p_1 p_2 m^2-i m_1 m(m_1^2-\mathbf{p}^2)}{(m_1^2-\mathbf{p}^2)^2} &
\frac{m^2 (m_1^2-p_1^2)}{(m_1^2-\mathbf{p}^2)^2}
\end{array}
\right),
\ee
where
\be
\rho=1 -\alpha t^2 \frac{p_2^2}{(m_1^2-\mathbf{p}^2)},
\ee
with eigenvalues
\bq
\lambda_1 &=& 0,  \\
\lambda_2 &=& 0,  \\
\lambda_3 &=& (1 + \beta t^2) \left(1+\frac{m^2(m_1^2+p_2^2)+ p_1^2(2 \tilde m^2-m^2)}{(m_1^2-\mathbf{p}^2)^2}\right).
\eq
In the equation above, we have
\be
\tilde m^2=(1+\beta t^2)m^2-\frac{\beta^2 t^4}{(1+\beta t^2)}p_2^2.
\ee
Again, we confirm that in a situation with a tiny Lorentz symmetry-breaking the MCS-like model respects, at tree level, unitarity and causality.

\subsubsection{$b_a$ timelike}

We repeat here the choice for $b^a$ used in the analysis of the self-dual model, for which the residue in the pole $m_1'^2$ of the propagator will yield only one non-zero eigenvalue, given by
\be
\lambda_3= 1 + \frac{m^2(m_1'^2+\mathbf{p}^2)}{(m_1'^2-\mathbf{p}^2)^2}.
\ee

We, thus, observe that the self-dual and the Maxwell-Chern-Simons models are physically equivalent. Besides, both models are stable and causal for the small Lorentz-symmetry violation, that is, in all concordant frames \cite{Kost-Lehnert}.

We note that there is a way to estimate the Lorentz-breaking parameters in the theory. Actually, in our theory the key parameter is $\beta b_ab_b=c_{ab}$. For small values of $\beta$ and $b_a$, one has $\alpha=\frac{\beta}{1-\beta b^2}\simeq \beta$ and, thus, one has $\alpha b_ab_b\simeq c_{ab}$. To find the value of $c_{ab}$, one can follow the way proposed in \cite{Kost-Lehnert}, finally arriving at that $|c_{ab}|\simeq \frac{m}{M_P}$, where $m$ is the mass in our theories and $M_P$ is the Planck mass.

\section{Concluding remarks}

We studied CPT-even extended versions of the 3D self-dual (SD) and Maxwell-Chern-Simons (MCS) models which violate Lorentz symmetry. This violation is accomplished by the addition of a Lorentz-breaking mass term to the self-dual model. The corresponding MCS-like Lagrangian density was obtained by means of the gauge embedding procedure. The duality was confirmed through the study of the two equations of motion, both in free and coupled cases, since it was found a mapping between the two vector fields $f_a$ and $A_a$, together with a mapping between the currents. The dualized model involves a Thirring-like current-current interaction and a magnetic coupling, as in \cite{Anacleto}.

In the sequel, we have shown that there exists a master action which generates the two models. A further investigation was carried out to check the equivalence of the two spectra. The massive poles for both theories are shown to coincide and to respect the requirements for unitarity and causality at tree level for controlled Lorentz-symmetry violation. Although the MCS-like model includes new poles, these new massless excitations have been shown to be restricted to non-dynamical sectors of the propagator.

Also, we note that within the dual mapping, the CPT-even extension of the SD theory is mapped to the CPT-even extension of the MCS theory, while, as it was showed in \cite{Fur}, the CPT-odd extension of the SD theory is mapped into the CPT-odd extension of the MCS theory. This situation differs from the four-dimensional case \cite{Wo} where the CPT-odd CFJ term is mapped into the CPT-even aether term. However, this difference seems to be essentially related with the dimensionality of the spacetime. To close the paper, we suppose that more sophisticated Lorentz-breaking extensions of duality are also possible.

\section{Appendix}

The Lorentz algebra of the wave operators used in the calculation of the propagators is shown in Table 1:
\begin{center}
\begin{tabular}{|c|c|c|c|c|c|c|c|c|c|c|}
\hline
& $\theta$ & $\omega$ & $S$ & $\Lambda$ & $\Sigma$ & $\Sigma^T$ & $A$ & $A^T$ & $B$ & $B^T$ \\
\hline
$\theta$ & $\theta$ & $0$ & $S$ & $\Lambda-\frac{\lambda }{\Box }\Sigma^T$ & $\Sigma-\lambda \omega$ & $0$ & $A$ & $A^T-\frac{\lambda}{\Box}B^T$ & $B$ & $0$ \\
\hline
$\omega$ & $0$ & $\omega$ & $0$ & $\frac{\lambda}{\Box }\Sigma^T$ & $\lambda \omega$ & $\Sigma^T$ & $0$ & $\frac{\lambda}{\Box}B^T$ & $0$ & $B^T$ \\
\hline
$S$ & $S$ & $0$ & $-\Box \theta$ & $A$ & $B$ & $0$ & $-\Box \Lambda+\lambda \Sigma^T$ & $C$ & $\Box\left(\lambda \omega -\Sigma\right)$ & $0$ \\
\hline
$\Lambda$ & $\Lambda-\frac{\lambda }{\Box}\Sigma$ & $\frac{\lambda }{\Box}\Sigma$ &$-A^T$ & $b^{2}\Lambda$ & $b^{2}\Sigma$ & $\lambda \Lambda$ & $0$ & $b^2 A^T$ & $0$ & $\lambda A^T$ \\
\hline
$\Sigma$ & $0$ & $\Sigma$ & $0$ & $\lambda \Lambda$ & $\lambda \Sigma$ & $\Box \Lambda$ & $0$ & $\lambda A^T$ & $0$ & $\Box A^T$ \\
\hline
$\Sigma^T$ & $\Sigma^T -\lambda \omega$ & $\lambda \omega$ & $-B^T$ & $b^2 \Sigma^T$ & $b^2 \Box \omega$ & $\lambda \Sigma^T$ & $0$ & $b^2 B^T$ & $0$ & $\lambda B^T$ \\
\hline
$A$ & $A -\frac{\lambda}{\Box} B$ & $\frac{\lambda}{\Box} B$ & $-C$ & $b^2 A$ & $b^2 B$ & $\lambda A$ & $0$ & $b^2 C$ & $0$ & $\lambda C$ \\
\hline
$A^T$ & $A^T$ & $0$ & $\Box \Lambda-\lambda \Sigma$ & $0$ & $0$ & $0$ & $\left(b^2 \Box -\lambda^2\right)\Lambda$ & $0$ & $\left(b^2 \Box -\lambda^2\right)\Sigma$ & $0$ \\
\hline
$B$ & $0$ & $B$ & $0$ & $\lambda A$ & $\lambda B$ & $\Box A$ & $0$ & $\lambda C$ & $0$ & $\Box C$  \\
\hline
$B^T$ & $B^T$ & $0$ & $\Box\left(\Sigma^T-\lambda \omega \right)$ & $0$ & $0$ & $0$ & $\left(b^2 \Box -\lambda^2 \right)\Sigma^T$ & $0$ & $\left(b^2 \Box -\lambda^2 \right)\Box \omega$ & $0$  \\
\hline
\end{tabular}
\vspace{0.3mm}

Table 1: Multiplicative table fulfilled by $\theta$, $\omega$, $S$, $\Lambda$,
$\Sigma$, $\Sigma^T$, $A$, $A^T$, $B$ and $B^T$. The products are supposed to obey the order ``row times
column''.
\end{center}

In the table above, we have
\be
C_{ab}= \left(b^2 \Box-\lambda^2 \right)\theta_{ab} - \lambda^2 \omega_{ab} - \Box \Lambda_{ab} + \lambda \left( \Sigma_{ab}+\Sigma_{ba}\right).
\ee

{\bf Acknowledgements.} This work was partially supported by Conselho
Nacional de Desenvolvimento Cient\'{\i}fico e Tecnol\'{o}gico (CNPq).
A. Yu. P. has been supported by the CNPq project 303438-2012/6.

\end{document}